\documentclass[twocolumn,preprintnumbers,amsmath,amssymb,aps]{revtex4}
\usepackage{graphicx}
\usepackage{dcolumn}
\usepackage[a4paper]{geometry}
\usepackage{xcolor}

\usepackage{bm}

\begin{document}

\title{Mode-locked Bloch oscillations in a ring cavity}

\author{M. Samoylova$^{a}$}
\author{N. Piovella$^{a}$}
\author{D. Hunter$^{b}$}
\author{G.R.M. Robb$^{b}$}
\author{R. Bachelard$^{c}$}
\author{Ph.W. Courteille$^{c}$}
\affiliation{$^{a}$Dipartimento di Fisica, Universit\`a degli Studi di Milano, via Celoria 16, I-20133 Milano, Italy\\
$^{b}$SUPA and Department of Physics, University of Strathclyde, John Anderson Building, 107 Rottenrow, Glasgow, G4 0NG, UK\\
$^{c}$Instituto de F\'isica de S\~ao Carlos, Universidade de S\~ao Paulo, 13560-970 S\~ao Carlos, SP, Brazil}

\date{\today}

\begin{abstract}
\textbf{Abstract}. We present a new technique for stabilizing and monitoring Bloch oscillations of ultracold atoms in an optical lattice under the action of a constant external force. In the proposed scheme, the atoms also interact with a unidirectionally pumped optical ring cavity whose one arm is collinear with the optical lattice. For weak collective coupling, Bloch oscillations dominate over the collective atomic recoil lasing instability and develop a synchronized regime in which the atoms periodically exchange momentum with the cavity field.
\end{abstract}

\maketitle

\section{Introduction}

Measuring the motional state of an atomic cloud generally requires the irradiation of a light beam and the detection of the velocity-dependent response of the cloud in the scattered light. Frequently used techniques either map the atomic distribution after a free expansion time, or measure the Doppler shift of the scattered light. However, the incident light also exerts optical forces which, in the case of ultracold atoms, can dramatically alter the atomic velocity and falsify its
measurement. A way to control the optical forces consists in making the light scattering process coherent, e.g.,~by forcing the scattered light into a single predefined light mode. The mechanical impact of the incident light then becomes predictable and can be taken into account, while heating can be avoided. Techniques based on this idea that have been successfully used in the past are the spectroscopy of recoil-induced resonances (RIR) \cite{Courtois1994,Kruse2003} or the collective atomic recoil laser (CARL) \cite{Bonifacio1994,Kruse2003b}. Both techniques allow to deduce the atomic velocity comparing the Doppler-shifted frequency of a single scattered light mode with the one of the incident light. In the case of CARL, the incident and the scattered light modes are the counterpropagating modes of a unidirectionally pumped ring cavity. Theoretical models and experiments have shown that the backaction of the atoms onto the light fields due to the CARL mechanism not only accelerates the atoms in a predictable way, but also provides an \emph{accurate} and \emph{continuous} monitor for the instantaneous atomic velocity \cite{VonCube2006,Slama2007}.

A particularly interesting application of techniques allowing continuous monitoring of the atomic velocity is the observation of Bloch oscillations of ultracold atoms stored in a one-dimensional optical lattice and subject to a constant external force, for instance, gravity \cite{BenDahan1996}. Since the Bloch oscillation frequency is directly proportional to the force, the observation of Bloch oscillations has become a fundamental tool for high precision gravity measurements \cite{Clade2005,Ferrari2006}. Recent proposals promise a continuous monitoring of the Bloch oscillation dynamics in a symmetrically pumped optical ring cavity in a way to minimize atomic backaction onto the amplitude and phase of the light fields leaking out of the cavity \cite{Peden2009,Goldwin2014}.

\begin{figure}[!ht]
    \centerline{\includegraphics[width=5.2cm]{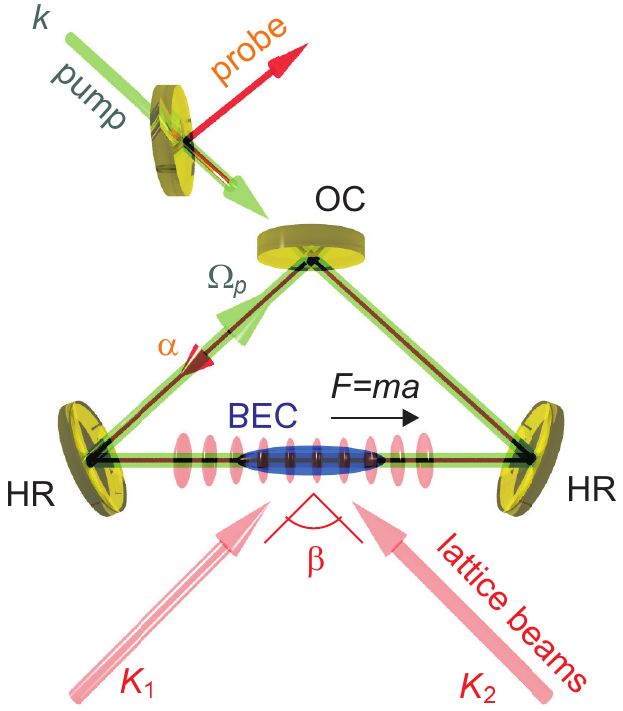}}
    \caption{(color online) Scheme of a ring cavity consisting of two high-reflecting mirrors (HR)
        and one output coupler (OC) interacting with a Bose-Einstein condensate (BEC) stored in one
        arm of the ring cavity. Only one cavity mode is pumped ($\Omega_p, k$), the counterpropagating probe
        mode ($\alpha$) is populated by backscattering from the atoms. Two lasers ($K_{1,2}$) crossing the
        cavity mode at the location of the BEC under angles $\pm\beta/2$ generate an optical lattice whose
        periodicity is commensurate with the standing wave created by the pump and probe modes. The atoms 
		are subject to an external accelerating force $F$.}
    \label{fig:fig1}
\end{figure}

In this letter, we analyze the CARL dynamics of ultracold atoms placed in a unidirectionally laser-pumped ring cavity in the presence of an \emph{externally imposed} 1D optical lattice aligned along the cavity axis. In addition, a constant force accelerating the atoms along the same axis is added to the system (see Fig.\ref{fig:fig1}). While the force incites the atoms to undergo Bloch oscillations in the imposed lattice, the CARL mechanism coherently scatters the pump light into the reverse mode in a self-amplified way accompanied by an atomic redistribution into a \emph{self-determined} 1D optical lattice, which competes with the externally imposed one.
When the cooperative coupling is stronger than the optical lattice strength, the CARL mechanism dominates over the Bloch oscillation dynamics and the population transfer between adjacent momentum states no longer occurs at the Bloch period, but depends on the CARL characteristic time. Unlike the resonance crossings in the adiabatic rapid passage (ARP) regime \cite{Peik1997}, the mechanism responsible for the momentum transition in this case is the scattering by the self-generated density grating.

Even though the CARL effect may strongly modify the Bloch oscillation frequency, for moderate cooperative coupling strength, we find a parameter range where the CARL and Bloch dynamics cooperate and synchronize giving rise to regular and stable Bloch oscillations. This is achieved through a cavity-mediated mode-locking mechanism between adjacent momentum modes. We find that the mode-locking is steady against technical noise, accidental excitations of atoms to higher bands, and dephasing due to interatomic interaction \cite{in preparation}. We investigate the transient regime between the two dynamics and derive the conditions under which pure Bloch oscillations can be observed.

In the following, we derive a model describing the interplay between CARL and Bloch oscillations and illustrate the mode-locking effect with numerical simulations.

\section{Model}

We consider a cloud of ultracold atoms confined in an optical standing wave with the lattice constant $\pi/k_l$. This standing wave can, for instance, be generated by two laser beams sufficiently far blue-detuned from an atomic resonance and intersecting at the location of the atoms under an angle $\beta$ given by $K\sin(\beta/2)=k_l$, where $K$ is the wavenumber of the laser beams, as shown in Fig.\ref{fig:fig1}. If the potential depth is denoted by $\hbar W_0$, the imposed potential reads as $(\hbar W_0/2)\sin(2k_lx)$. Being additionally exposed to the force potential $max$, with the atomic mass $m$ and the acceleration $a$, the atoms execute Bloch oscillations with frequency $\nu_b=ma/2\hbar k_l$.

We now add an optical ring cavity, letting the atoms simultaneously interact with its two counterpropagating cavity modes with wavenumber $k_0$. The atomic motion in such an environment has been experimentally shown \cite{Kruse2003,Nagorny2003,Slama2007} to act back onto the intracavity light fields and imprint into their phases and amplitudes detectable signatures. In certain parameter regimes this backaction, known as CARL \cite{Bonifacio1994}, develops self-synchronized atomic trajectories in conjunction with the spontaneous formation of a standing wave optical potential. It is thus reasonable to expect observable signatures of atomic Bloch oscillations when the externally imposed standing wave is commensurate with the standing wave formed by the ring cavity, i.e.,~$k_0=k_l$.

With $d$ being the electric dipole moment of the atomic transition and $E_1$ the electric field generated by a single photon in a cavity mode, the atom-field coupling strength is $\Omega_1=dE_1/\hbar$. The Rabi frequency generated by the pump light is $\Omega_p$, and $\Delta$ is its detuning from the atomic resonance. Thus, we can express the atom-mediated pump-probe coupling strength as $U_0=\Omega_1\Omega_p/4\Delta$. Labeling the probe mode as $\alpha$, where $|\alpha|^2$ is the photon number, the interference between pump and probe modes generates a dipolar potential with the depth $\hbar\alpha|U_0|$ along the optical axis of the ring cavity. Starting from the basic equations describing the model \cite{Gatelli2001} and disregarding atomic interaction in sufficiently dilute atomic clouds, we can write the equations of motion for the probe mode $\alpha$ and the atomic wave function $\psi$ in the following form:
\begin{align}
    i\hbar\tfrac{\partial\psi}{\partial t} & = -\tfrac{\hbar^2}{2m}\tfrac{\partial^2\psi}{\partial x^2}
			-i\hbar U_0\left(\alpha e^{2ik_0x}-\alpha^*e^{-2ik_0x}\right)\psi \nonumber\\
		& -max\psi+\hbar\tfrac{W_0}{2}\sin(2k_0x)\psi, \label{eqn:CARL_BEC1}\\
        \tfrac{d\alpha}{dt} & = NU_0\int|\psi|^2e^{-2ik_0x}d(2k_0x)+(i\delta-\kappa)\alpha, \label{eqn:CARL_BEC2}
\end{align}
where $N$ is the number of atoms, $\kappa$ is the cavity decay width, and $\delta=\omega_0-\omega_s$ is the pump-probe detuning. For $W_0=0$ and $a=0$, the equations describe the usual CARL dynamics \cite{Gatelli2001}.

The evolution of the system can be conveniently described in the accelerated frame moving with a momentum $mat$ along the positive direction of the $x$-axis. In this frame, the wave function is transformed according to $\psi=\tilde\psi\exp(imaxt/\hbar)$. Substituting $\alpha=\tilde\alpha-\alpha_0$ with $\alpha_0=W_0/4U_0$ into Eq.(\ref{eqn:CARL_BEC1}) and Eq.(\ref{eqn:CARL_BEC2}), we obtain:
\begin{align}
    \tfrac{\partial\tilde\psi}{\partial t} & = \tfrac{i\hbar}{2m}\left(\tfrac{\partial}{\partial x}+\tfrac{imat}{\hbar}\right)^2\tilde\psi\nonumber\\
        & -U_0\left(\tilde\alpha e^{2ik_0x}-\tilde\alpha^*e^{-2ik_0x}\right)\tilde\psi, \label{eqn:CARL_BEC3}\\
    \tfrac{d\tilde\alpha}{d t} & = NU_0\int|\tilde\psi|^2e^{-2ik_0x}d(2k_0x)\nonumber\\
        & +(i\delta-\kappa)(\tilde\alpha-\alpha_0). \label{eqn:CARL_BEC4}
\end{align}
This shows that the impact of the externally imposed standing wave can be simply accounted for as an additional laser beam pumping the probe mode at the rate $\alpha_0\kappa$.

The size of the atomic cloud is assumed to be much longer than the radiation wavelength and its density is uniform. Thus we can expand the atomic wave function into plane waves with periodicity $\pi/k_0$, i.e., $\tilde\psi(x)=\tfrac{1}{\sqrt{2\pi}}\sum_nC_n(t)e^{2ink_0x}$, where $|C_n|^2$ is the probability of finding the atoms in the $n$th momentum state. Note that the wavefunction is expanded in the momentum states $|2\hbar k_0\cdot n\rangle$~\cite{Peik1997} rather than the extensively used Bloch states $|n_b,q\rangle$ with the quasi-momentum $q$ and the band index $n_b$ ~\cite{Gluck2002}. Using the above definition of the Bloch oscillation frequency and introducing the single-photon recoil frequency $\omega_r=\hbar k_0^2/2m$, we obtain:
\begin{align}
    \tfrac{d C_n}{d t} & = -4i\omega_r(n+\nu_bt)^2C_n+U_0\left(\tilde\alpha^*C_{n+1}-\tilde\alpha C_{n-1}\right), \label{eq:C_n}\\
    \tfrac{d\tilde\alpha}{d t} & = U_0N\sum\limits_nC_{n-1}^*C_n+(i\delta-\kappa)(\tilde\alpha-\alpha_0). \label{eq:alpha}
\end{align}
We now assume the cavity decay to be much faster than the Bloch or CARL dynamics, such that $\kappa\tilde\alpha\gg d\tilde\alpha/dt$, and the detuning to be small on the scale of the cavity linewidth, i.e., $\delta\ll\kappa$. In this regime, the cavity can be adiabatically eliminated, resulting in a light field which is slaved to the collective atomic motion:
\begin{equation}
	\tilde\alpha \approx \alpha_0+\frac{NU_0}{\kappa}\sum\limits_nC_{n-1}^*C_n. \label{adiab}
\end{equation}
The last term in Eq.(\ref{adiab}) represents the backaction of the atoms onto the cavity field. The type of dynamics described by Eq.(\ref{eq:C_n}) and (\ref{adiab}) depends critically on the cooperative coupling of the atoms to the cavity fields, which can be controlled via the number of atoms $N$. For $NU_0/\kappa\ll\alpha_0$, the cooperative coupling is very weak, so that the atomic backaction onto the cavity fields may be disregarded. The cavity field decouples from the atoms and quickly evolves into a steady state given by $U_0\tilde\alpha=U_0\alpha_0=W_0/4$. In this case, we recover the usual Bloch oscillation picture, where the motion of the atoms is governed by Eq.(\ref{eq:C_n}) and can be interpreted as follows \cite{Peik1997}: In the frame accelerated by the external force, the frequencies of the two counterpropagating light fields are Doppler-shifted, and the effect of the external force manifests itself as a linear chirp in the first term on the right-hand side of the equation. As time goes on, a resonance is crossed at $t=-n\tau_b$, where $\tau_b=1/\nu_b$ is the Bloch period, and the crossing is periodically repeated for every $n=-1,-2,\dots$. At every crossing, if the ARP condition $16\nu_b/\omega_r\ll(W_0/\omega_r)^2\le 1$ is fulfilled, the atoms change their momentum by $2\hbar k$, transferring one photon from one beam of the optical lattice to the other one. This momentum transfer causes a force which compensates for the external force in the laboratory frame. In an equivalent picture, the accelerated atomic matter wave decreases its de Broglie wavelength until, at the edges of the Brillouin zone, it becomes commensurate with the optical lattice and is Bragg-reflected.

For larger cooperative coupling, $NU_0/\kappa$, the ring cavity comes into play. Now, the matter wave may not only scatter light between the optical lattice beams, but it also cooperatively scatters photons from the pumped cavity mode into the reverse mode $\alpha$, which now exert influence on the atomic dynamics. If $NU_0/\kappa\gg\alpha_0$, the CARL mechanism dominates over the Bragg scattering. In this regime, the mechanism responsible for transferring momentum to the atoms is not the ARP across the Bragg resonance, but the backscattering of the pump light by a self-generated atomic density grating \cite{Bonifacio1994}. As a consequence, the population transfer between adjacent momentum states does not occur at the regular Bloch periods, but may vary in time.

Only for moderate cooperative coupling ($NU_0/\kappa\approx\alpha_0$), we find a parameter range where CARL and Bragg scattering cooperate to set up a synchronized regime with regular and stable Bloch oscillations. 
At some point, when the backscattering of the pump light into the probe mode becomes stronger, the depth of the potential formed in the cavity by interference of the pump and the counterpropagating probe light may exceed the depth of the optical lattice generated by the external beams $K_1$ and $K_2$. In this regime, the CARL mechanism takes over and imposes its dynamics on the atoms~\cite{Bonifacio1994,Kruse2003,Slama2007}, dominating the Bloch oscillations.

\section{Simulations}

\begin{figure}[!ht]
    \centerline{\includegraphics[width=6.9cm]{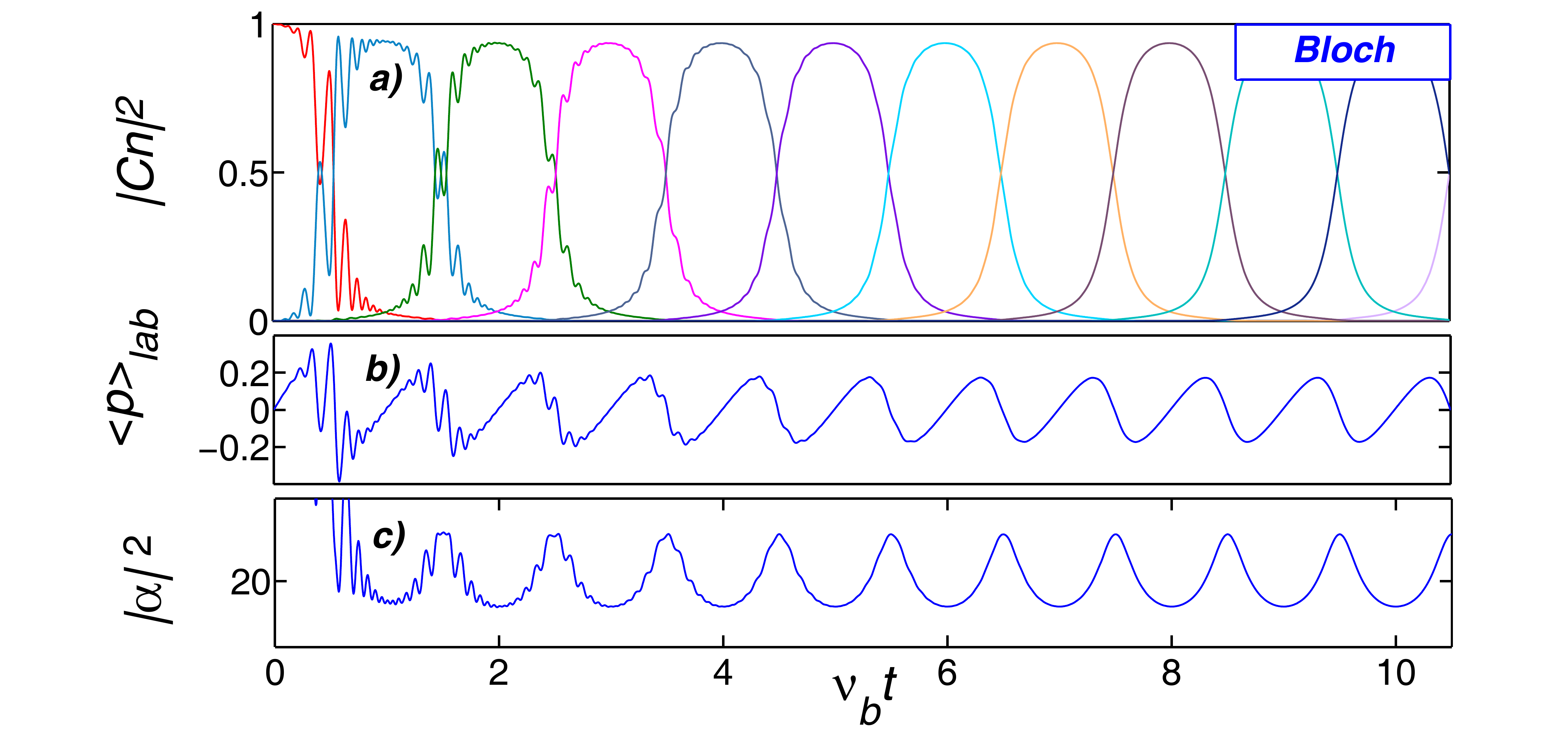}}
    \caption{(color online) Time evolution of (a) the population of the momentum states $|C_n|^2$, 
		(b) the average atomic momentum $\langle p\rangle_{lab}$ in the laboratory frame with $N=4\cdot10^4$ atoms, and 
		(c) the average photon number $|\alpha|^2$ in the regime dominated by Bloch dynamics. The parameters used to perform the 
		simulations are: $\alpha_0=20$, $\nu_b=0.035\omega_r$, $\kappa=160\omega_r$, $\delta=0$, and $U_0=0.04\omega_r$.}
    \label{fig:fig2}
\end{figure}

\begin{figure}[!ht]
    \centerline{\includegraphics[width=6.9cm]{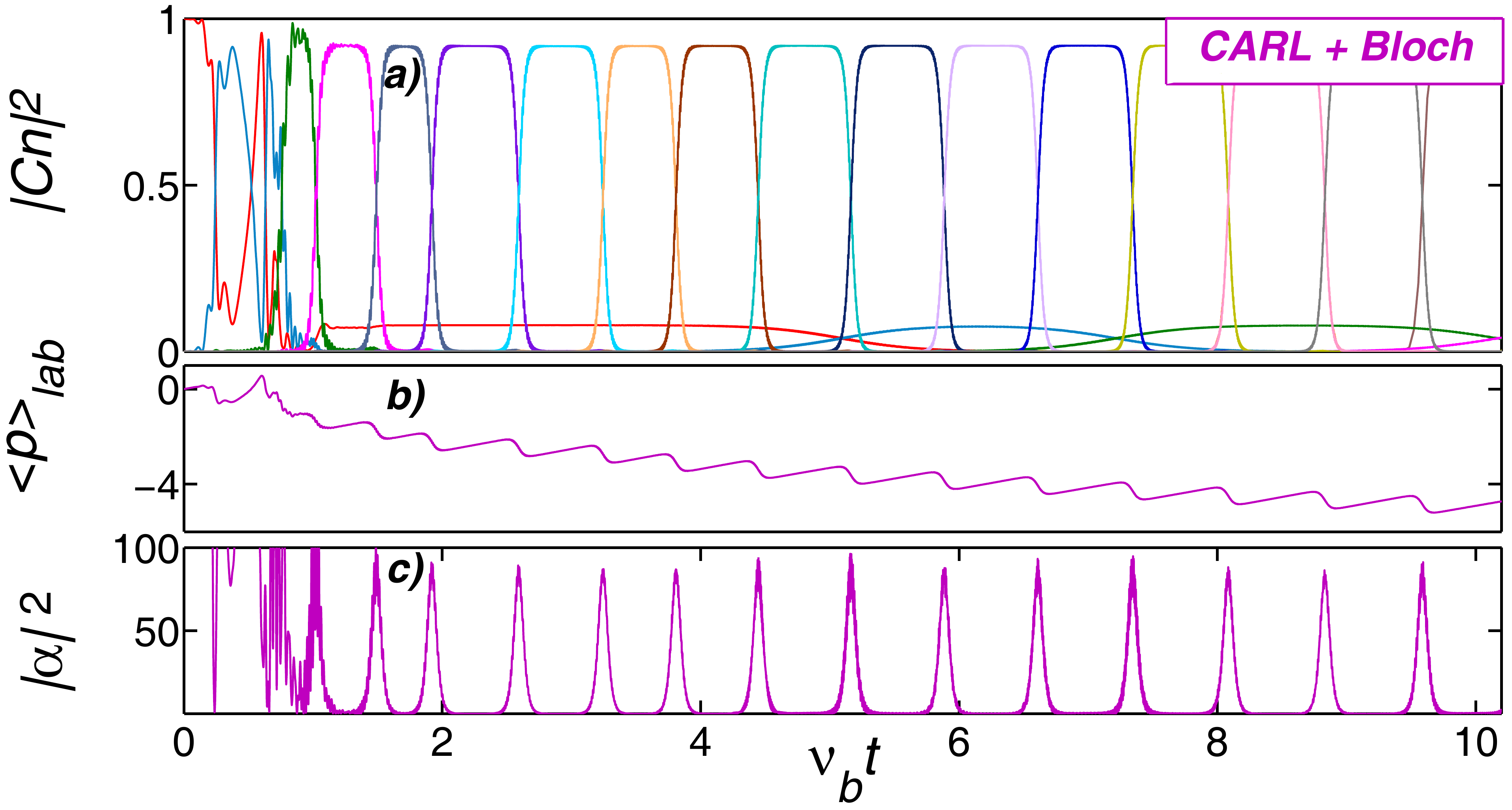}}
    \caption{(color online) Dynamics of (a) the momentum state populations, 
		(b) the average atomic momentum in the laboratory frame, and 
		(c) the average photon number as a function of normalized time $\nu_bt$ in a regime, where CARL and Bloch dynamics compete.
		$N=8\cdot10^4$ and the other parameters are the same as in Fig.\ref{fig:fig2}.}
    \label{fig:fig3}
\end{figure}

\begin{figure}[!ht]
    \centerline{\includegraphics[width=6.9cm]{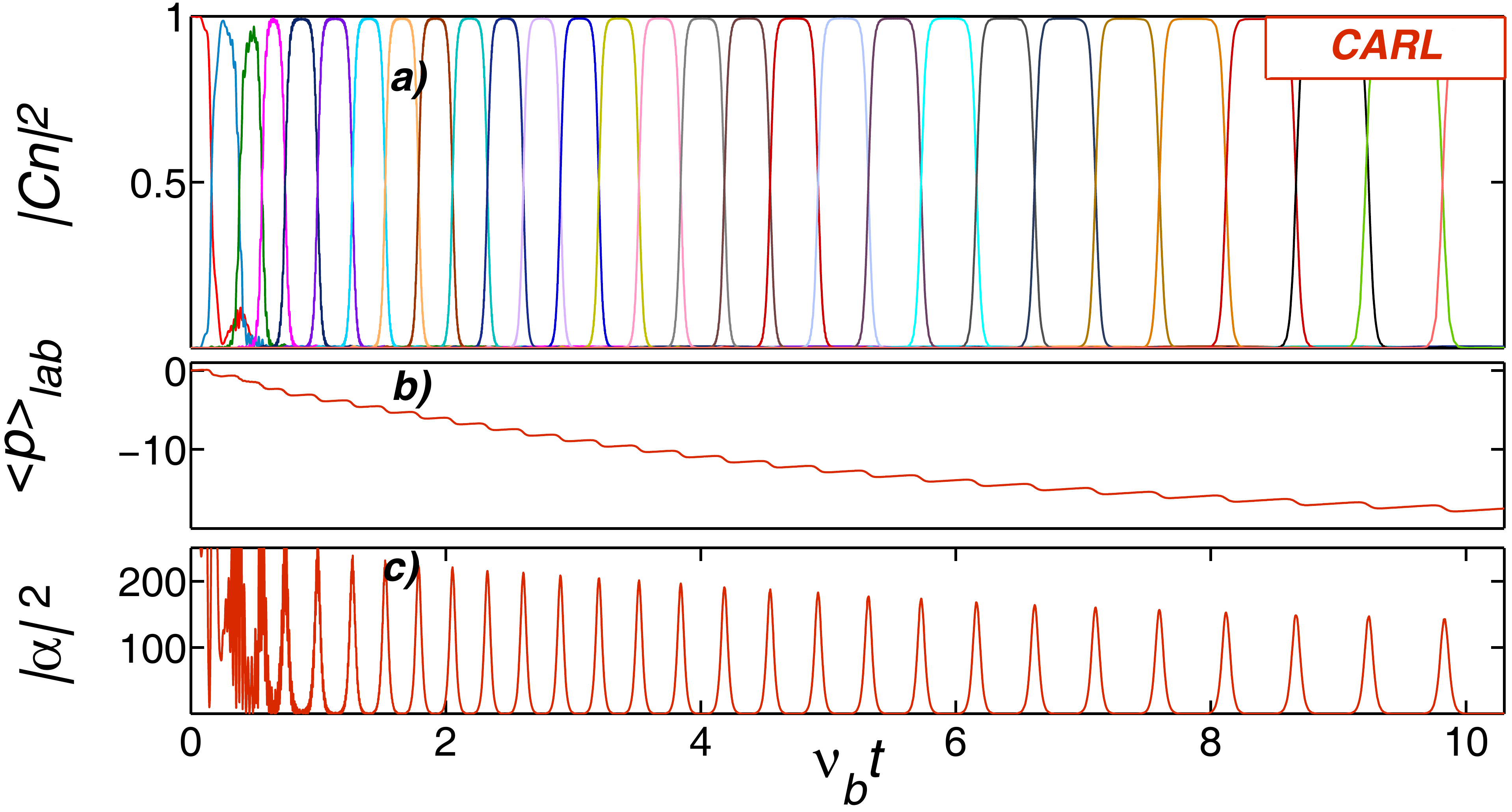}}
    \caption{(color online)  Dynamics of (a) the momentum state populations, 
		(b) the average atomic momentum in the laboratory frame, and 
		(c) the average photon number as a function of normalized time $\nu_bt$ in a regime, dominated by CARL dynamics.
		$N=12\cdot10^4$ and the other parameters are the same as in Fig.\ref{fig:fig2}.}
    \label{fig:fig4}
\end{figure}

Figs.\ref{fig:fig2}-\ref{fig:fig4} illustrate the intricate dynamics in the regimes dominated by Bloch oscillations or by CARL dynamics, as well as an intermediate regime where both dynamics compete. We choose the example of an ultracold cloud of $^{87}$Rb atoms interacting with the light fields via their D2-line at $\lambda_0=780~$nm, for which the recoil frequency is $\omega_r=(2\pi)~3.75~$kHz and the Bloch oscillation frequency, supposing that the accelerating force is gravity (i.e.~$a=g$) is $\nu_b=0.035\omega_r$. We also assume $\kappa=160\omega_r$, $\delta=0$, $U_0=0.04\omega_r$, and $W_0=3.2\omega_r$, which corresponds to $|\alpha_0|^2=400$ photons. The collective coupling strength is controlled by varying the atom number between $N=4\cdot10^4$ and $12\cdot10^4$. These parameters are perfectly realizable in state-of-the-art experiments.

Fig.\ref{fig:fig2} represents a regime dominated by Bloch oscillation dynamics. Fig.\ref{fig:fig2}(a) shows a typical evolution of the momentum state populations $|C_n|^2$  as a function of scaled time $\nu_b t$ for the case that the dynamics is dominated by Bloch oscillations. The population of each momentum state is accentuated by a different color in order to facilitate their visual distinction during the temporal evolution. As can be seen, all atoms initially prepared in a single momentum state participate in the dynamics. This is explained by the fact that throughout the evolution the momentum transfer between adjacent momentum states remains fully efficient. As a consequence, the Bloch oscillations persist for long times, as seen in Fig.\ref{fig:fig2}(b) showing the evolution of the average atomic momentum in the laboratory frame, $\langle p\rangle_{\textrm{lab}}=\langle p\rangle+\nu_b t$ with $\langle p\rangle=\sum_n n|C_n|^2$. After a transient of approximately three Bloch oscillations the population is efficiently restored into the first Brillouin zone and the feedback provided by the cavity field onto the atomic motion tends to assist the adiabatic rapid passages between momentum states helping to complete the momentum transfer each Bloch period $\tau_b$. 

Moreover, the atomic Bloch oscillation dynamics is accompanied by a radiation field reaching, after a transient, a stationary regime characterized by periodic bursts of light emitted into the probe mode at each oscillation. The intracavity photon number evolution $|\alpha|^2$ in the probe mode is demonstrated in Fig.\ref{fig:fig2}(c). The average photon number $|\alpha|^2\simeq20$ corresponds, for the chosen value of $\kappa$, to a photon flux of $\sim18400~\text{s}^{-1}$ outside the cavity behind the output coupler, i.e., $\sim140$ photons/Bloch oscillation. Hence, the light bursts appear to be perfectly detectable via a photon counter and to provide a reliable and stable monitor of the atomic motion.

In the intermediate regime, when both dynamics are present, only a fraction of the atoms perform Bloch oscillations, whereas the remaining atoms fail to synchronize. This is  illustrated in Fig.\ref{fig:fig3}(a). The competition between CARL dynamics and Bloch oscillations leads to irregular oscillation frequencies, and the dispersion of the atoms over different momentum states induces drifts of the average atomic momentum [see Fig.\ref{fig:fig3}(b)]. Moreover, the bursts of light in the radiation field shown in Fig.\ref{fig:fig3}(c) are no longer periodic and cannot be used as a reliable signature of the atomic dynamics.

In contrast to the previous cases, in the regime dominated by CARL dynamics, the atoms quickly jump from one momentum state to the next one in a superradiant fashion. This can be seen in Fig.\ref{fig:fig4}(a). The backscattering of the pump light and the amplification of a coherent wave in the probe mode of the ring cavity is accompanied by an increasing drift of the average atomic momentum [see Fig.\ref{fig:fig4}(b)]. At longer times the increase slows down because the Doppler shift associated with the atomic motion drives the scattered light out of the cavity resonance, and as a result the CARL force diminishes. In this regime, the radiation field experiences multiple light bursts per Bloch oscillation period [see Fig.\ref{fig:fig4}(c)] and, consequently, its dynamics cannot be considered as a reliable monitor of the atomic motion.

\begin{figure}[!ht]
    \centerline{\includegraphics[width=6.9cm]{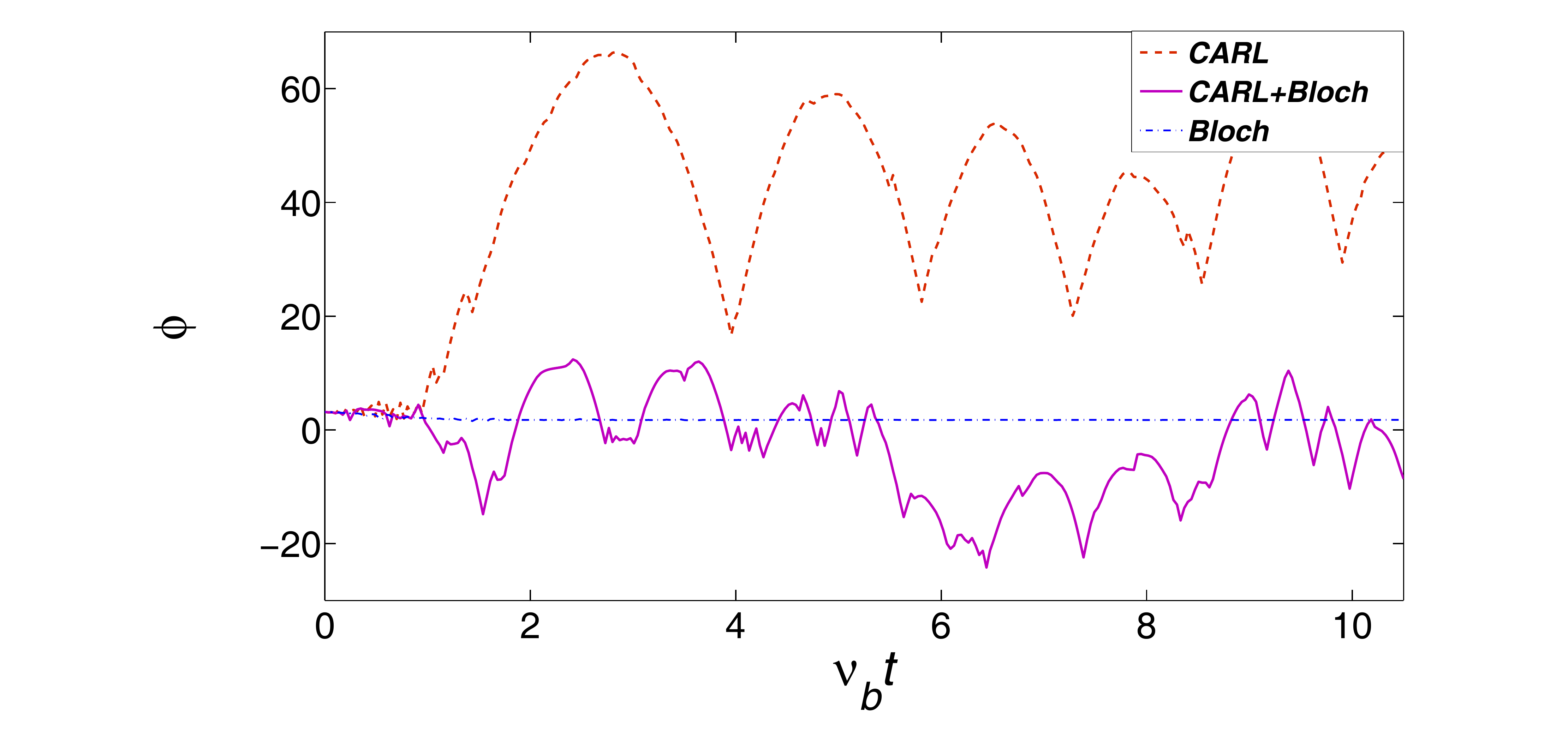}}
    \caption{(color online) Phase $\phi$ of the probe field $\alpha=\tilde{\alpha}-\alpha_0$ in 
		the CARL regime with $N=12\cdot10^4$ (red), the intermediate regime with $N=8\cdot10^4$ (purple), 
		and the Bloch regime with $N=4\cdot10^4$ (blue) as a function of scaled time $\nu_b t$. 
		The remaining parameters are the same as in Fig.\ref{fig:fig2}.}
    \label{fig:LPLPhase}
\end{figure}

The transition between the regimes is illustrated by Fig.\ref{fig:LPLPhase}, which shows the time evolution of the phase $\phi$ of the probe field $\alpha=\tilde{\alpha}-\alpha_0$ in the regime of pure Bloch oscillations (blue), as well as in the intermediate regime (purple) and the regime dominated by CARL (red). Being absolutely unstable in the case of CARL dynamics, the phase fluctuations gradually decrease in the intermediate regime, when the Bloch dynamics becomes stronger. And they totally disappear when the Bloch oscillations dominate. Then the feedback provided by the cavity stabilizes the Bloch oscillations, and after some transient rapid oscillations (which could be reduced by an adiabatic switch-on of the lattice potential), the phase remains locked to $\pi/2$ with only slight perturbations at each Bloch oscillation.

\section{Conclusion}

The atom-field coupling in a uni-directionally pumped ring cavity provides a feedback mechanism of the atomic motion onto the amplitude and phase of the counterpropagating light field. This feedback mechanism, named CARL, can be exploited to stabilize and monitor Bloch oscillations of ultracold atoms in an optical lattice. Whether CARL and the Bloch oscillation dynamics perturb each other or cooperate and synchronize, depends on the collective atom-field coupling strength. We observed, for a certain range of parameters, a mode-locking of the atomic motion to the Bloch oscillation frequency, which leads to regular and stable Bloch oscillations. The fact that the atoms are not accelerated, but stay within the first Bloch band, is at the origin of several important practical advantages. Firstly, long-term drifts of the atomic center-of-mass motion are prevented. Moreover, robust light bursts emitted into the probe mode provide a non-destructive monitor of the Bloch oscillations over long times. Note that, if the pump laser is tuned close to the atomic resonance, the intracavity light field $\Omega_p$ can exert a non-negligible constant radiation pressure on the atoms, which may alter the measure of the atomic acceleration and, hence, the frequency of the Bloch oscillations. It has, however, no impact on the mode-locking dynamics.

\bigskip\noindent\textbf{\emph{Acknowledgements.\---~}} We are grateful to J. Goldwin for stimulating discussions. This work was funded by the Research
Executive Agency (Program COSCALI, Grant No. PIRSES- GA-2010-268717) and the Funda\c{c}\~ao de Amparo \`a  Pesquisa do Estado de S\~ao Paulo (FAPESP).
M.S. acknowledges the support from the internship program of the Instituto de F\'isica de S\~ao Carlos (IFSC).

\end{document}